# New models of quantum games

Paulo Benício Melo de Sousa *Rubens Viana Ramos and José Tarcísio Costa Filho

*Department of Teleinformatic Engineering, Federal University of Ceara, Campus do Pici, 710, C.P. 6007, 60755-640, Fortaleza-Ceará, Brazil*

**Abstract**

In this work we propose and develop modified quantum games (zero and non-zero sum) in which payoffs and strategies are entangled. For the games studied, Nash and Pareto equilibriums are always obtained indicating that there are some interesting cases where quantum games can be applied.

*PACS:* 03.67.Dd

*Keywords:* Quantum games, entanglement, quantum circuits

Quantum games were introduced by Meyer [1] and Eisert [2] as generalization of classical games [3,4]. Some interesting implementations of quantum games based on the use of entanglement have been proposed [5-7], even thought some authors consider that it does not represent a requirement for quantum games [8]. Most quantum games proposed are versions of their classical counterpart as, for instance, PQ penny flip [1], quantum battle of sexes [9], quantum Monty Hall problem [10], quantum market games [11], quantum Parrondo games [12] and, the most famous, the quantum prisoner's dilemma [2-6,12,13]. Basically, in a game, two (or more) players are trying to reach a final state which ensures them the best benefit, acting individually or cooperating with other players. The common representation of a game consists of a set of rational $k$ players ($k \geq 2$) and a allowed set of strategies $\left(\{s\}_i^k\right)$, each strategy profile implying in a payoff value ($\$_k$).

___________________
*Corresponding author
*E-mail addresses*: benicio@deti.ufc.br, rubens@deti.ufc.br, tarcisio@deti.ufc.br

The players choose the strategies aiming to obtain the best payoff value. The most discussed game, the two prisoners' dilemma, has important features: non-cooperative environment, non-zero sum characteristic and simultaneous actions. It can be described as follows: two persons, accused to commit a crime, have to choose between two strategies: cooperate (*C*) or to defect (*D*) with police. The payoffs are the years that will be discounted from their penalty. The usual payoff values for this game are shown in Table 1.

Table 1. Payoffs for prisioner's dilemma.

| Alice \ Bob | C | D |
|---|---|---|
| C | (3,3) | (0,5) |
| D | (5,0) | (1,1) |

The goal for each player is to maximize his/her individual payoff. The best for players *A* and *B* is to choose the strategies $s_A^*$ and $s_B^*$ such that $\$_A(s_A^*, s_B^*) \geq \$_A(s_A, s_B^*), \forall s_A \in S_A$ and $\$_B(s_A^*, s_B^*) \geq \$_B(s_A^*, s_B), \forall s_B \in S_B$. This correspond the action of defecting for both. This situation implies the Nash equilibrium. There is no rational assumption in which they can get an optimal equilibrium, the Pareto optimal, in which both would cooperate and, together, they could reach the best payoff for both (3,3).

The classical prisoner dilemma game can be seen as a physical system having the initial state |*CC*⟩ and the strategies correspond to choose between identity (*I*, cooperate) and NOT ($\sigma_x$, defect) operators. Hence, the possible final states are {|*CC*⟩,|*CD*⟩,|*DC*⟩,|*DD*⟩} and, after a measurement, one can always know the strategies chosen. The quantum prisoners' dilemma game is a direct generalization of that scenario. Any single qubit quantum gate can be used (and not only *I* and $\sigma_x$) and entangled states

are allowed. The general scheme is shown in Fig. 1. The gate $J$ is the entangler gate, $U_A$ and $U_B$ are the individual strategies for $A$ and $B$ players, respectively; $J^\dagger$ is the adjoint gate responsible for disentanglement operation and, at last, $M_A$ and $M_B$ are measurers.

In the quantum circuit of Fig. 1, $U_A$ and $U_B$ are general single-qubit gates given by:

$$U_{A(B)}(\theta,\phi,\varphi) = \begin{pmatrix} e^{-i\phi_{A(B)}} \cos\frac{\theta_{A(B)}}{2} & -e^{i\varphi_{A(B)}} \sin\frac{\theta_{A(B)}}{2} \\ e^{-i\varphi_{A(B)}} \sin\frac{\theta_{A(B)}}{2} & e^{i\phi_{A(B)}} \cos\frac{\theta_{A(B)}}{2} \end{pmatrix}. \qquad (1)$$

In (1), $\theta \in [0,\pi]$ and $\phi,\varphi \in [-\pi,\pi]$. It has been proved that using $U(\theta,\phi=0,\varphi=0)$, the quantum game is equivalent to a classical game with mixed strategies [4]. The expression for the entangler operator $J$ is [4,11,13]:

$$J = e^{i\frac{\gamma}{2}\sigma_x \otimes \sigma_x}. \qquad (2)$$

One easily gets that $J|CC\rangle = \cos(\gamma/2)|CC\rangle + i\sin(\gamma/2)|DD\rangle$. Hence, the parameter $\gamma$, belonging to the interval $[0, \pi/2]$, controls the amount of entanglement of the state. The gate $J$ can be decomposed as $J = CNOT \cdot \exp(i\gamma\sigma_x/2) \otimes \sigma_0 \cdot CNOT$ and, therefore, the quantum circuit for the quantum prisoner's dilemma is as shown in Fig. 2.

The game is played varying $U_A$ and $U_B$, while $\gamma$ is a fixed structural parameter of the game. The output state of the game is $|\Psi\rangle = J^{\dagger}(U_A \otimes U_B)J|CC\rangle$ and the average payoff for each player is given by:

$$\langle \$_i \rangle_Q = \$_{CC,i}|\langle\Psi|CC\rangle|^2 + \$_{CD,i}|\langle\Psi|CD\rangle|^2 + \$_{DC,i}|\langle\Psi|DC\rangle|^2 + \$_{DD,i}|\langle\Psi|DD\rangle|^2, \qquad (3)$$

where $i = A, B$ and, for example, $\$_{CD,A}$, from Table 1, is 0, while $\$_{CD,B}$, from Table 1, is 5. On the other hand, the classical game using mixture of strategies can be described by the probabilities of cooperate and defect of each player. Let us assume that $r(q)$ is the probability of player $A(B)$ to cooperate. In this case, the average payoff for each player is:

$$\langle \$_i \rangle_C = \$_{CC,i} rq + \$_{CD,i} r(1-q) + \$_{DC,i}(1-r)q + \$_{DD,i}(1-r)(1-q) \qquad (4)$$

It is interesting to analyze under what conditions $\langle\$_i\rangle_Q = \langle\$_i\rangle_C$ occurs. Firstly, we stress that the condition $\langle\$_i\rangle_Q = \langle\$_i\rangle_C$ does not imply that quantum and classical games are equivalent in a formal way, since the number and the type of equilibriums are not been considered. However, one can argue if it is advantageous to play a quantum game if, in average, he/she could win the same playing a simpler classical game. Firstly, let us suppose that, given a payoff table for the Prisoners Dilemma, the best choice for players $A$ and $B$ are the strategies $U_A$ and $U_B$, giving the average payoffs $\langle\$_A\rangle_Q$ and $\langle\$_B\rangle_Q$. Then, it is possible to find a mixed classical game such that $\langle\$_i\rangle_C = \langle\$_i\rangle_Q$ if

$$\max\left(\frac{\langle \$_B \rangle_Q - \$_{DD,B}}{\langle \$_A \rangle_Q - \$_{DD,A}}, \frac{\alpha_B}{\alpha_A}\right) \leq \frac{\gamma_B}{\gamma_A} \leq \frac{\beta_B}{\beta_A} \tag{5}$$

$$\alpha_{A(B)} = \$_{CD,A(B)} - \$_{DD,A(B)} \tag{6}$$

$$\beta_{A(B)} = \$_{DC,A(B)} - \$_{DD,A(B)} \tag{7}$$

$$\gamma_{A(B)} = \$_{CC,A(B)} - \$_{CD,A(B)} - \$_{DC,A(B)} + \$_{DD,A(B)} \tag{8}$$

An interesting point of the quantum game is the fact that the strategies can not be known from the results of the measurement of the output state, in other words, it is not possible to identify a possible traitor, what maybe an advantage in some real situations.

Aiming to advance in the structure of quantum games, we propose in this work a new model of quantum games that considers the payoffs as quantum information too. In this case, one can have entanglement between the payoffs and the individual strategies. As will be seen latter, this entanglement between payoffs and strategies can lead the game to remain only in equilibriums states or, seen from another point of view, one can make a game for which some possibilities of the payoff table are forbidden. A quantum circuit for the Prisoners Dilemma having strategies and payoffs entangled is shown in Fig. 3.

In order to show the usefulness of the circuit of Fig. 3, let us to analyze the particular case for which $J|CC\rangle=(|CC\rangle+i|DD\rangle)/2^{1/2}$ ($\gamma=\pi/2$), and $U_A$ and $U_B$ are chosen from the set $\{I, \sigma_x\}$, that is, the game structure is quantum but both players are classical. The codification of the payoff qubits (the two upper qubits) are as follows: $|00\rangle\rightarrow(1,1)$, $|01\rangle\rightarrow(0,5)$, $|10\rangle\rightarrow(5,0)$ and $|11\rangle\rightarrow(3,3)$. The four possible output states are:

$$|\Phi\rangle_{II} = \frac{1}{\sqrt{2}}\left[|00\rangle\left(\frac{|CC\rangle+i|DD\rangle}{\sqrt{2}}\right)+|11\rangle\left(\frac{|CC\rangle-i|DD\rangle}{\sqrt{2}}\right)\right] \quad (9)$$

$$|\Phi\rangle_{I\sigma_x} = |11\rangle|CD\rangle \quad (10)$$

$$|\Phi\rangle_{\sigma_x I} = |11\rangle|DC\rangle \quad (11)$$

$$|\Phi\rangle_{\sigma_x\sigma_x} = \frac{1}{\sqrt{2}}\left[|00\rangle\left(\frac{|CC\rangle+i|DD\rangle}{\sqrt{2}}\right)-|11\rangle\left(\frac{|CC\rangle-i|DD\rangle}{\sqrt{2}}\right)\right] \quad (12)$$

where, for example, $|\Phi\rangle_{\sigma_x I}$ means the output state when $U_A=\sigma_x$ and $U_B=I$. Observing (9)-(12) one easily sees that only Nash equilibrium and Pareto optimal are obtained. No classical game can have a similar behavior. If *A* and *B* players choose their strategies (cooperate and do not cooperate) with the same probability, then the Pareto optimal is obtained with probability 0.75 while the Nash equilibrium happens with probability 0.25. This new model of Prisoners Dilemma game can also be thought as a regulator mechanism avoiding the monopoly in a duopoly system. That is, the situation where only one prisoner (or an enterprise in an economic context) wins, never happens. Hence, the situations (5,0) and (0,5) in Table 1 are forbidden by the game structure. However, non-cooperative results are still possible avoiding unconditional cooperation. Another direct application is from the perspective of computational resources. Let us suppose that two softwares are running in parallel in a computer and both of them are requiring CPU time. The game proposed, that can be implemented by an operational system, avoids situations in which the CPU is dedicated to only one software, stopping the other.

Other important class of games evolves zero-sum payoffs. Differently from the previous game presented, each player gets what other (or others, considering more than two players) loses. In this case, the payoff matrix is, for instance, as shown in the Table 2 (where both players can adopt strategies S1 or S2). Note in Table 2 that $\langle \$_A \rangle + \langle \$_B \rangle = 0$.

Table 2 – Possible Table of payoffs for a zero-sum game.

| Alice \ Bob | S1 | S2 |
|---|---|---|
| S1 | (0,0) | (-2,2) |
| S2 | (1,-1) | (0,0) |

A possible quantum physical implementation of a zero-sum quantum game is presented in Fig. 4, in which the players $A$ and $B$, locally distant, use two pure bipartite entangled states. Initially, player $A$ enters with qubits $U_A|0\rangle \otimes (|0\rangle+|1\rangle)/2^{1/2} = (a|0\rangle+b|1\rangle) \otimes (|0\rangle+|1\rangle)/2^{1/2}$ and player $B$ enters with qubits $U_B|0\rangle \otimes (|0\rangle+|1\rangle)/2^{1/2} = (c|0\rangle+d|1\rangle) \otimes (|0\rangle+|1\rangle)/2^{1/2}$. The choices of $U_A$ and $U_B$ are, respectively, $A$'s and $B$' strategies. The unitary matrix $U_{AB}$ is a four qubit operation providing as output two qubits for player $A$ and two qubits for player $B$. The unitary operator $U_{AB}$ works as follows:

$$U_{AB}\left[a|0\rangle+b|1\rangle\right]_A \otimes \left[\frac{|0\rangle+|1\rangle}{\sqrt{2}}\right]_A \otimes \left[c|0\rangle+d|1\rangle\right]_B \otimes \left[\frac{|0\rangle+|1\rangle}{\sqrt{2}}\right]_B = |\Psi\rangle \quad (13)$$

$$|\Psi\rangle = \left(a|00\rangle+b|11\rangle\right)_{13} \otimes \left(c|00\rangle+d|11\rangle\right)_{24} = \left[ac|0000\rangle+bc|1010\rangle+ad|0101\rangle+bd|1111\rangle\right]_{1234} \quad (14)$$

From (14) one realizes that $A$ and $B$ share two entangled states. The two first qubits (1 and 2) are measured by $A$ using the measurers $M_1^A$ and $M_2^A$, respectively, while and the last two qubits (3 and 4) are measured by $B$ using the measurers $M_1^B$ and $M_2^B$, respectively. Considering these information, a simple game can be implemented using the following rules: player $A$ wins (which implies that $B$ loses) if the results of the measurements in $M_1^A$ and $M_2^A$ are equal (the same happens with the results in $B$'s measurers). Obviously, $B$ wins the game if the values measured by $M_1^B$ and $M_2^B$ are different (the same happen with the results of $A$'s detectors). The probabilities of $A$ and $B$ to win are, hence, given by:

$$P_A = 1 - \left(|a|^2 - 2|a|^2|c|^2 + |c|^2\right) \quad (15)$$
$$P_B = 1 - P_A \quad (16)$$

If one consider a classical game in which $A$ chooses $U_A=I$ with probability $p$ and $U_A=\sigma_x$ with probability $(1-p)$, and $B$ chooses $U_B=I$ with probability $q$ and $U_B=\sigma_x$ with probability $(1-q)$, then, the probability of $A$ win the game is $p_A=p-2pq+q$, while $B$ wins with probability $(1-p_A)$. Hence, in this game, one can find classical game as good as the quantum version. The Nash equilibrium occurs when $P_A=P_B=0.5$. Now, one can change the quantum game in order to include quantum payoffs in the zero-sum game, creating an entanglement between each player strategies and his/her payoff. In this case, the circuit for the quantum game can be as shown in Fig. 5.

In Fig. 5, the relations between the states of the output payoffs qubits and costs in Table 2 are $\{|00\rangle \to (0,0), |01\rangle \to (-1,1), |10\rangle \to (2,-2), |11\rangle \to (0,0)\}$. The evolution of the states in the quantum circuit is as follows:

$$|\Psi_1\rangle = |000000\rangle_{123456} \qquad (17)$$

$$|\Psi_2\rangle = |00\rangle_{12} \otimes (a|0\rangle + b|1\rangle)_3 \otimes \left(\frac{|0\rangle+|1\rangle}{\sqrt{2}}\right)_4 \otimes (c|0\rangle + d|1\rangle)_5 \otimes \left(\frac{|0\rangle+|1\rangle}{\sqrt{2}}\right)_6 \qquad (18)$$

$$|\Psi_3\rangle = |00\rangle_{12} \otimes U_{AB}\left[(a|0\rangle + b|1\rangle)_3 \otimes \left(\frac{|0\rangle+|1\rangle}{\sqrt{2}}\right)_4 \otimes (c|0\rangle + d|1\rangle)_5 \otimes \left(\frac{|0\rangle+|1\rangle}{\sqrt{2}}\right)_6\right] =$$

$$= |00\rangle_{12} \otimes [ac|0000\rangle + bc|1010\rangle + ad|0101\rangle + bd|1111\rangle]_{3456} \qquad (19)$$

$$|\Psi_4\rangle = ac|000000\rangle + bc|111010\rangle + ad|000101\rangle + bd|111111\rangle =$$

$$= \underbrace{|00\rangle_{12}(ac|0000\rangle + ad|0101\rangle)_{3456}}_{Nash\ Equilibrium} + \underbrace{|11\rangle_{12}(bc|1010\rangle + bd|1111\rangle)_{3456}}_{Nash\ Equilibrium} \qquad (20)$$

Hence, as can be observed in (20), the Nash equilibrium is always obtained. In this kind of game, the players never lose and never win too. Obviously one can think what is the importance of a game in which one never wins.

Summarizing, the use of entanglement between strategies and payoffs can increase the probabilities to obtain Nash and Pareto equilibriums, as well, it can permit the construction of games for which some situations are forbidden, what can be a desirable characteristic in some cases, as for example, resources sharing. For the cases studied, Pareto and Nash equilibriums are always achieved. The results using the Prisoner Dilemma game were especially relevant, since we have obtained a better situation in comparison to the classical game. On the other hand, for the zero-sum game proposed, only the Nash equilibrium is obtained evidencing that there is no equilibrium better than a non-profit payoff.

**Acknowledgements**

This work was supported by the Brazilian agency FUNCAP.

# FIGURE 1

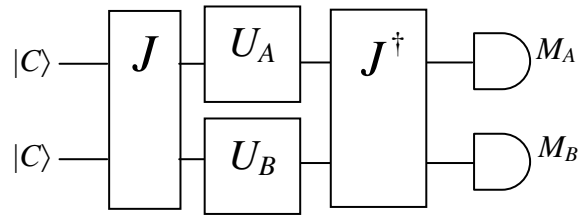

Fig. 1. Quantum circuit for quantum prisoners' dilemma.

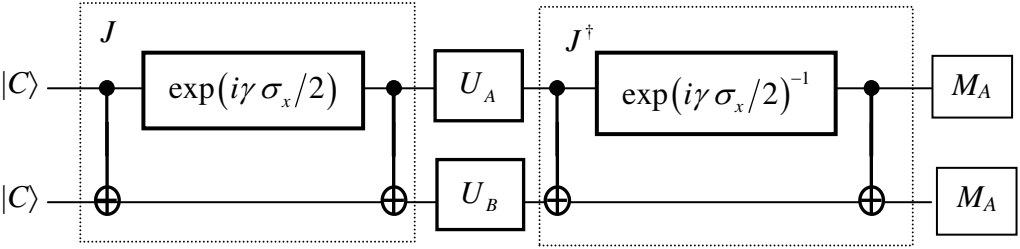

Fig. 2. Quantum circuit for quantum prisoners' dilemma game using only CNOTs and single qubit gates.

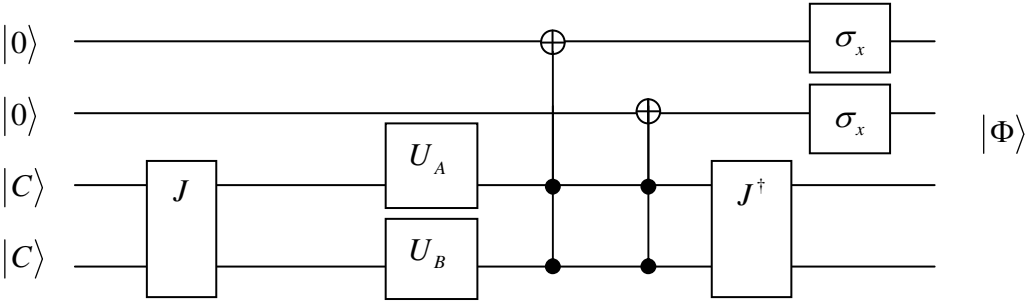

Fig. 3. Quantum prisoners' dilemma game having strategies and payoffs entangled.

# FIGURE 4

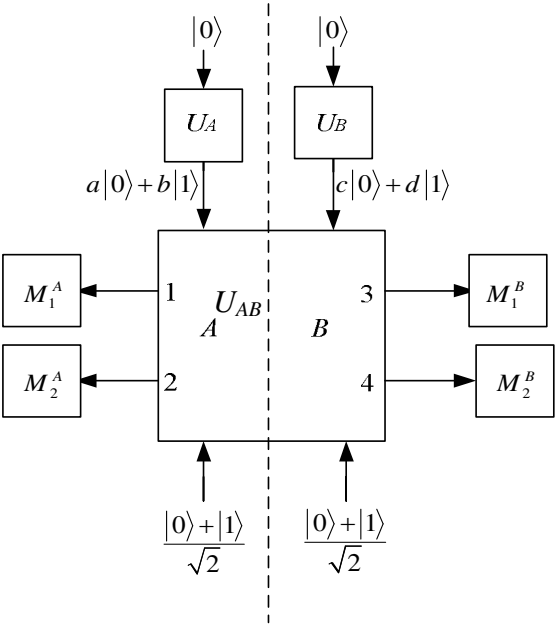

Fig. 4. Zero-sum quantum game.

# FIGURE 5

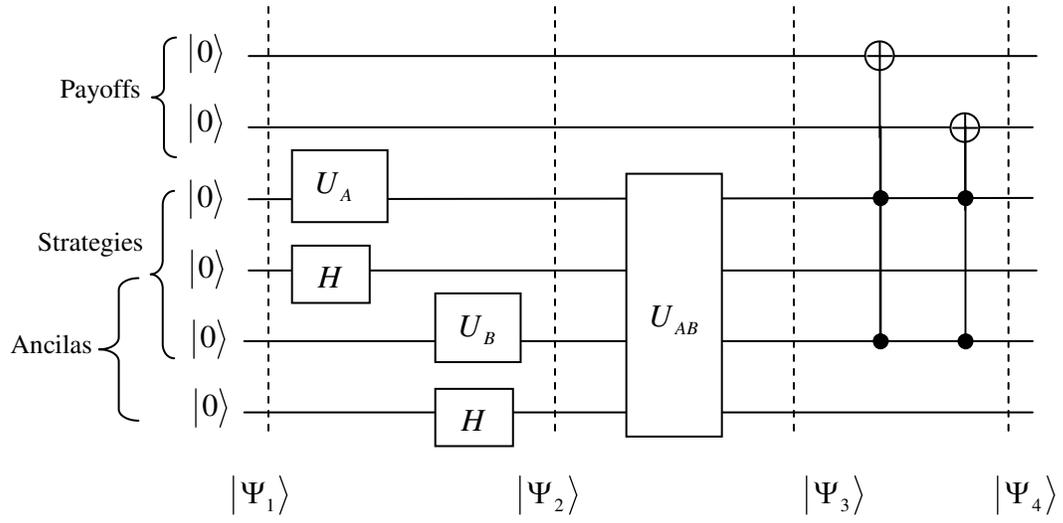

Fig. 5 – Quantum circuit for zero-sum game having entangled strategies and payoffs.